\documentclass[twocolumn,prb,amsmath,amssymb,amsfonts,superscriptaddress,floatfix,nopacs]{revtex4}
\usepackage{graphicx}
\usepackage{dcolumn}
\usepackage{bm}
\usepackage{rotating}
\usepackage{color}
\usepackage{amsmath}

\def\be{\begin{equation}}
\def\ee{\end{equation}}
\def\bd{\begin{displaymath}}
\def\ed{\end{displaymath}}
\def\-{\phantom{-}}
\usepackage{lipsum}

\begin{document}

\title{Evolution of magnetic Dirac bosons in a honeycomb lattice}

\date{\today}

\author{D. Boyko}
\affiliation{Department of Physics, University of North Florida, Jacksonville, FL 32224, USA}
\author{A. V. Balatsky}
\affiliation{Institute of Material Science, Los Alamos National Laboratory, Los Alamos, New Mexico 87545}
\affiliation{Nordic Institute for Theoretical Physics, KTH Royal Institute of Technology and Stockholm University, Roslagstullsbacken 23, 106 91 Stockholm, Sweden}
\author{J. T. Haraldsen}
\affiliation{Department of Physics, University of North Florida, Jacksonville, FL 32224, USA}

\begin{abstract}

We examine the presence and evolution of magnetic Dirac nodes in the Heisenberg honeycomb lattice. Using linear spin theory, we evaluate the collinear phase diagram as well as the change in the spin dynamics with various exchange interactions. We show that the ferromagnetic structure produces bosonic Dirac and Weyl points due to the competition between superexchange interactions. Furthermore, it is shown that the criteria for magnetic Dirac nodes are coupled to the magnetic structure and not the overall crystal symmetry, where the breaking of inversion symmetry greatly affects the antiferromagnetic configurations. The tunability of the nodal points through variation of the exchange parameters leads to the possibility of controlling Dirac symmetries through an external manipulation of the orbital interactions.

%We examine the spin waves and presence of Dirac nodes for various collinear magnetic phases of the Heisenberg honeycomb lattice. Using linear spin theory through an exact diagonalization approach, we analyze the effect of nearest, next-nearest, and next-next-nearest exchange couplings in each spin configuration and determine the general phase diagrams for the spin configurations. While the ferromagnetic and standard antiferromagnetic case are stable with zero anisotropy, the more complex antiferromagnetic phases require multiple exchange couplings as well as a critical anisotropy to become stable. Furthermore, the presence of multiple exchange couplings produce Dirac-like nodes in the magnon spectra for most structures. This most evident in the ferromagnetic configuration, where the addition of next and next-next nearest neighbor interactions creates two additional mode crossovers. We also show that criteria for magnetic Dirac nodes are coupled to magnetic structure and not the crystal symmetry, where the antiferromagnetic case loses its Dirac nature due to the breaking of inversion symmetry regardless of exchange coupling.

\end{abstract}

\maketitle

\section{Introduction}

In the past few years, Dirac and Weyl materials have gained much attention in the field of condensed matter physics due to their unique properties, primarily the relativistic transport of fermions in low-energy excitations.\cite{wehl:14,cays:13,Du:15,chen:16,ross:12,gara:13,gome:12,burk:11} Dirac materials provide an interesting symmetry protection; especially for 2D and topological materials,\cite{cays:13,xu:13} which provides the potential for technological applications through the tunability of the electronic interactions to external fields.\cite{li:13,lens:15,wang:15,chir:17}

The main criteria of a Dirac material is the presence of a Dirac cone, which is a multi-band crossover of at least two distinct modes in the electronic structure that typically occurs near the Fermi level and produces a nodal point called a Dirac node or point.\cite{wehl:14} Therefore, the system or structure in question must have a non-Bravias lattice (two sublattice (2SL) or greater). Furthermore, Dirac nodes have a four-fold degeneracy produced through a coupling of momentum to a spin or pseudospin, where the Dirac coupling provides a chirality in the lattice of a Dirac material, thus creating a positive and negative mode and a linear crossover at the Fermi level.\cite{wehl:14,cays:13} Here, the presence of a Dirac node inhibits direct backscattering of electrons by requiring the addition of a flip of the (pseudo)spin due to a real space inversion.\cite{wehl:14} However, if inversion symmetry is broken, then a gap can form, and the Dirac cone is destroyed. These two criteria lay down the general foundation for a Dirac material as a non-Bravias lattice in the presence of inversion symmetry, which is a challenge since inversion symmetry can be broken easily through structure distortions, impurities, and dopants, or the presence of external fields.

\begin{figure}
  \centering
  \includegraphics[width=2.25in]{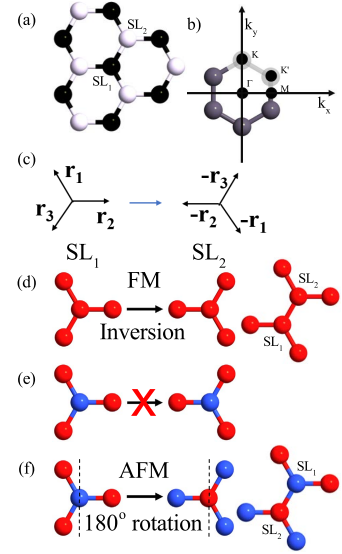}
  \caption{Illustrated honeycomb structure (a) and reciprocal space (b). The honeycomb structure has two sublattices (black and white) that are related through inversion symmetry (c). When magnetic structure is imposed, a ferromagnetic structure (d) maintains inversion symmetry. However, an antiferromagnetic configuration breaks inversion symmetry (e) and produces a $180^o$ rotation (f).}
  \label{fig:inversion}
\end{figure}

Furthermore, Dirac nodes should not be confused with Weyl nodes or points, where a Weyl point is typically described as a topological phenomenon that is essentially in the massless limit of the Dirac equation.\cite{wehl:14} Through the breaking of parity or time-reversal symmetry, the four-fold degeneracy of the Dirac point can split into two Weyl points,\cite{Du:15} which tends to make Weyl points more stable against perturbations since their degenerate crossover cannot be broken with changing symmetries\cite{chen:16}. Overall, Weyl points can be determined by the Chern number of the valence band, which requires that Weyl points occur in two at $k$ and -$k$ with equal Chern number.\cite{youn:12,mont:09} 
 
%which is shown in the Dirac Hamiltonian
%\begin{equation}
%    H(k)=v_F \bar k \cdot \bar \sigma,
%\end{equation}
%\noindent where $\bar k$ is the momentum vector, $\bar \sigma$ is the (pseudo)spin, and $v_F$ is the Fermi velocity. 

Graphene is the most well-known electronic Dirac material due to the non-Bravias honeycomb lattice that produces a two-sublattice (2-SL) structure with a real-space inversion (illustrated in Fig. \ref{fig:inversion}(a)) \cite{geim:07}. The honeycomb lattice produces four Dirac fermions, which consists of two spin-degenerate cones in each of the two valleys near symmetry points $K$ and $K^{\prime}$. This creates a total of six Dirac cones at the corners of the first Brillouin zone (Fig. \ref{fig:inversion}(b)),\cite{neto:09} where the nodes have Berry curvature\cite{xiao:07} and can shift in $k$-space depending on various parameters.\cite{tarr:12} 

As demonstrated above, the majority of the research on Dirac materials has been focused on the creation and evolution of Dirac fermions in the electronic structure with a few groups examining the presence of Dirac symmetries in the magnetic structure.\cite{fran:16,ower:16} Therefore, we set out to investigate the existence of Dirac bosons in hopes of providing further insights into the nature of Dirac symmetries and the possible realization of spintronic applications for materials with these properties.

In this study, we show the presence of magnetic Dirac and Weyl bosons in the ferromagnetic and anti-ferromagnetic honeycomb structures by examining the presence of mode crossovers in the spin-wave spectra. Using a Holstein-Primakoff expansion of the Heisenberg spin-spin exchange Hamiltonian, we determine the phase diagram and spin dynamics for various magnetic configurations and show that the development of Dirac magnons is dependent on the magnetic structure due to the breaking of inversion symmetry in certain phases. Furthermore, we examine the evolution of the magnetic Dirac and Weyl points with the addition of multiple exchange interactions. The presence of Weyl nodes in the magnetic structure is a product of frustration produced by next and next-next nearest neighbor interactions. Additionally, we show that a more complicated interaction system allows for the stability of at least three other collinear phases that produce various nodal points.

\begin{figure}
  \centering
  \includegraphics[width=3.25in]{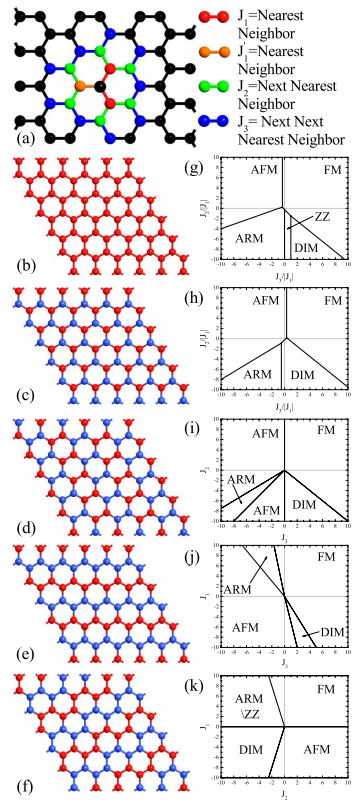}
  \caption{(a) An illustration of the exchange parameters with nearest neighbor interactions $J_1$ and $J_1^{\prime}$, next-nearest neighbor interactions $J_2$, and next-next-nearest neighbor interactions $J_3$. In most cases, $J_1$ = $J_1^{\prime}$. However, $J_1^{\prime}$ is used to explore the effects of asymmetry some phases. (b) A ferromagnetic (FM), (c) An anti-ferromagnetic (AFM), (d) zig-zag (ZZ), (e) dimerized (DIM), and (f) an armchair (ARM). Using the classical energy for each configuration, we can determine various phase diagrams for these configurations (f-j), where (g) $J_1=1$, (h) $J_1=-1$, (i) $J_1=0$, (j) $J_2=0$, and (k) $J_3=0$.}
  \label{fig:PhaseStructures}
  \end{figure}

\section{Magnetic Hamiltonian and Classical Energy}

To examine the creation of Dirac bosons in the magnetic structure, the starting point is to use a similar structure that is used for Dirac fermions. Therefore, one should be able to create a basic magnetic Dirac structure through the use of the honeycomb lattice. The most common collinear magnetic structures for the honeycomb lattice are ferromagnetic (FM), antiferromagnetic (AFM), zig-zag (ZZ), dimerized (DIM), and armchair (ARM) (illustrated in Fig. \ref{fig:PhaseStructures}). While there are most undoubtedly non-collinear phases in this structure, we consider only select collinear configurations that exist above the critical anisotropy points.
 
The introduction of spin into the honeycomb lattice creates a complication to the presence of inversion symmetry in the 2-SL structure. Shown in Fig. \ref{fig:inversion}(d) the FM configuration will maintain the inversion symmetry of the honeycomb lattice and will produce a Dirac cone. This has been discussed in Ref [\onlinecite{fran:16}]. However, in an antiferromagnetic configuration, the inversion symmetry is broken due to the spin-flip from SL$_1$ to SL$_2$ (Fig. \ref{fig:inversion}(e)). The antiferromagnetic configuration produces a $180^{\circ}$ rotation that makes SL$_1$ magnetically equivalent to SL$_2$. Therefore, it is expected that only a Goldstone mode will appear in the spin-wave spectra. Therefore, as the other configurations are examined, we expect to find further complication, since the appearance of magnetic Dirac nodes is contingent, not only on the structural inversion and sublattice but the magnetic degree of freedom as well.

\begin{figure}
  \centering
  \includegraphics[width=3.25in]{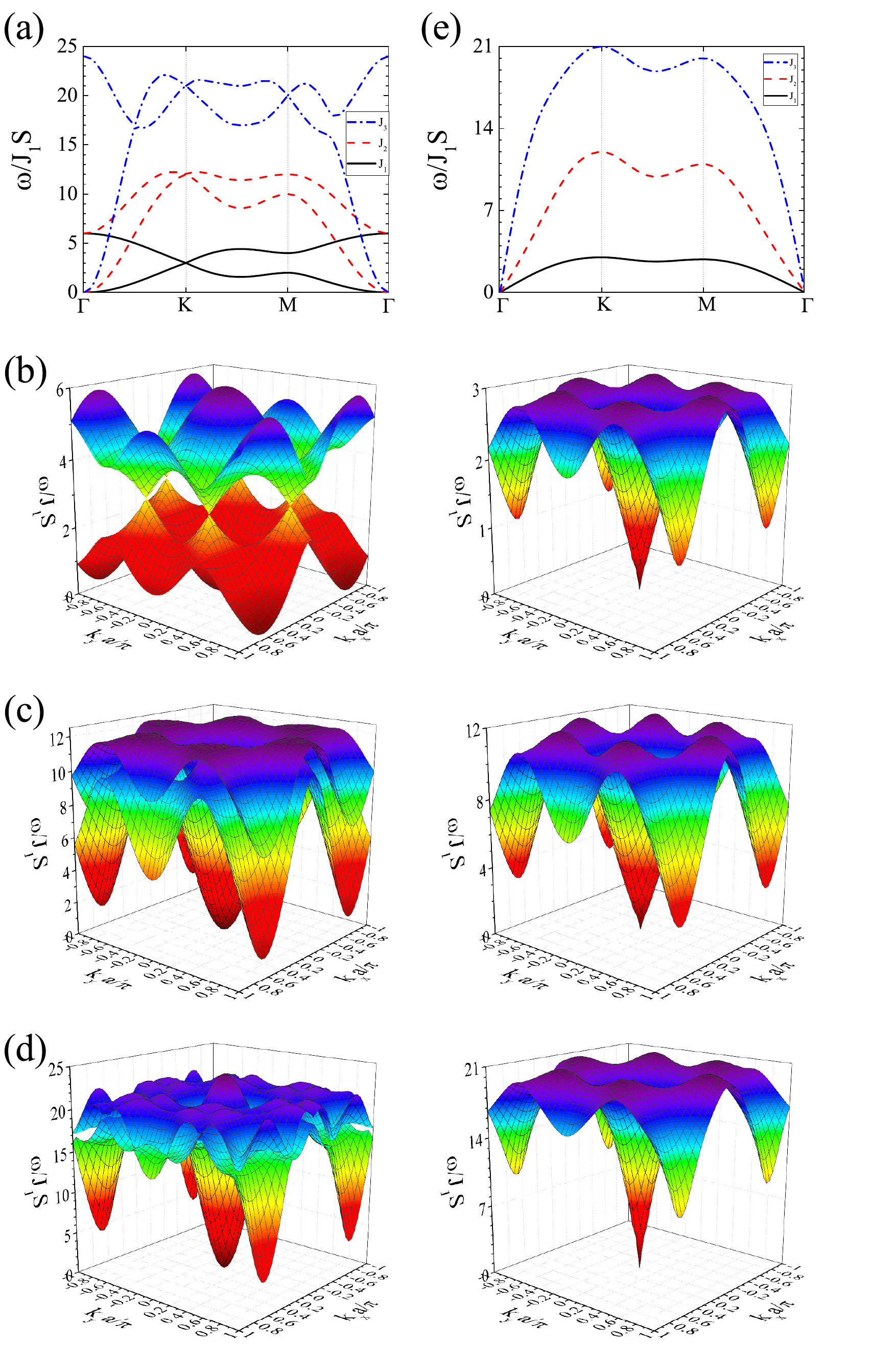}
  \caption{(a, b) The 2-D spin waves of a FM and AFM honeycomb structure with added levels of exchange-interactions, respectively. (c, d) The 3-D spin waves of the FM and AFM honeycomb structure with just a $J_1$ spin exchange interaction, respectively. (e, f) The 3-D spin waves of the FM and AFM honeycomb structures with $J_1$, and $J_2$ spin exchange interactions, respectively. (g, h) The 3-D spin waves of the FM and AFM honeycomb structures with $J_1$, $J_2$, and $J_3$ spin exchange interactions, respectively.}
  \label{fig:fmafm}
\end{figure}

To model the spin dynamics in the honeycomb lattice, nearest ($J_1$ and $J_1^{\prime}$), next nearest ($J_2$), and next-next nearest ($J_3$) neighbor interactions on the sublattices of the honeycomb lattice are considered (shown in Fig. \ref{fig:PhaseStructures}). Using a Holstein-Primakoff expansion of the Heisenberg spin-spin exchange Hamiltonian

\begin{equation} \label{Energy}
    H= -\frac{1}{2} \sum_{i \neq j} J_{ij} \bar S_i \cdot \bar S_j - D \sum_{i} S_{iz}^2,
\end{equation}

\noindent the energies and spin dynamics for the aforementioned configurations are determined\cite{kitt:87,hara:09JPCM}. Here, $J_{ij}$ are the super-exchange interactions between spins $\bar S_{i}$ and $\bar S_{j}$ at sites $i$ and $j$. $D$ is an anisotropy which will keep the collinear phases stable.\cite{hara:09JPCM}  $J>0$ denotes a FM exchange and $J<0$ is AFM. In most cases, $J_1$ = $J_1^{\prime}$. However, $J_1^{\prime}$ is used to explore the effects of asymmetry in some phases. The ZZ, DIM, and ARM phases are produced through a competition of exchange interactions due to the frustration introduced by $J_2$ and $J_3$. 

Through a $(1 / S)$ expansion, the nature of the various orders of this Hamiltonian can be shown as

\begin{equation} \label{hamexpansion}
    H = E_0 + H_1 + H_2 + \cdot \cdot \cdot,
\end{equation}

\noindent where $E_0$ gives the classical energy, $H_1$ is the vacuum contribution to the spin waves, and $H_2$ provides the spin dynamics. Since we are considering the semi-classical approximation, we may ignore higher-order terms due to quantum fluctuations being negligible at $T=0$ and large $S$.\cite{hara:09JPCM}

For each configuration, the classical energy can be written as, 

\begin{equation}\label{energies}
\begin{aligned}
     &E_{FM} & &= -\frac{1}{2}( 3 J_1 + 6 J_2 + 9 J_3 )\\
     &E_{AFM} & &= -\frac{1}{2}( - 3 J_1 + 6 J_2 - 9 J_3 )\\
     &E_{ZZ} & &= -\frac{1}{2}( J_1 - 2 J_2 - J_3 )\\
     &E_{DIM} & &= -\frac{1}{2}( - J_1 - 2 J_2 + J_3 )\\
     &E_{ARM} & &= -\frac{1}{2}( J_1 - 2 J_2 - 3 J_3 ).
\end{aligned}
\end{equation}

\begin{figure}
  \centering
  \includegraphics[width=3.0in]{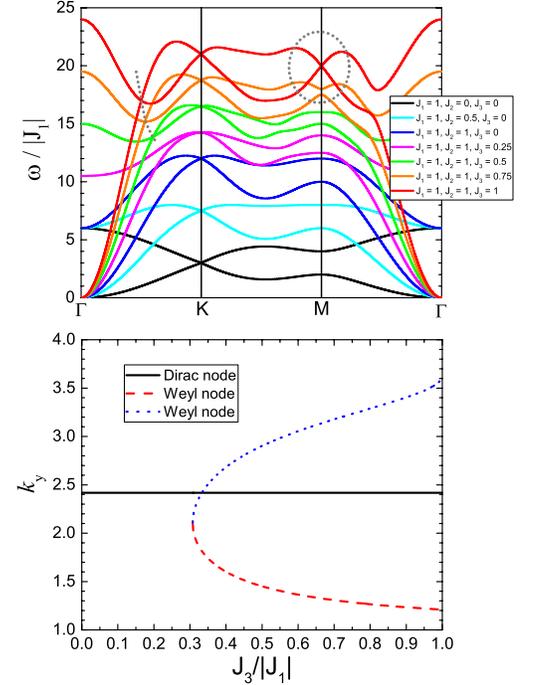}
  \caption{(a) Evolution of Dirac and Weyl nodes as next-next nearest neighbor interactions are introduced.(b) The creation of the two Weyl modes along $k_y$ with increasing $J_3$.}
  \label{fig:FM-evo}
\end{figure}

\begin{figure}
  \centering
  \includegraphics[width=3.25in]{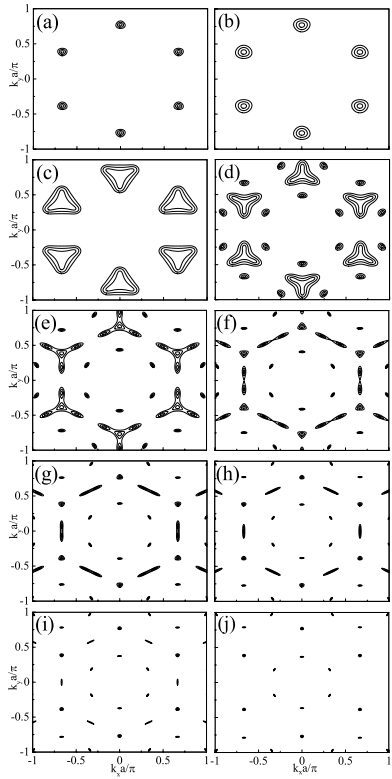}
  \caption{Evolution of Dirac and Weyl nodes as $J_3/|J_1|$ is increased. The panels show the nodal points for $J_1^{\prime}/|J_1|$ = 1, $J_2$ = 0, $J_3/|J_1|$ =  (a) 0, (b)  0.15, (c) 0.3, (d) 0.45, (e)  0.6, (f)  0.75, (g)  0.90, (h)  1.05, (i)  1.20, (j)  1.35.}
  \label{fig:J3e}
\end{figure}

\begin{figure}
  \centering
  \includegraphics[width=3.25in]{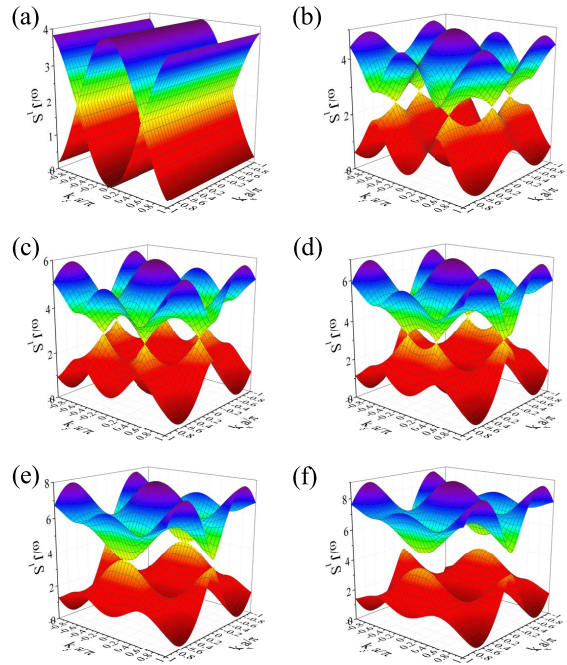}
  \caption{Evolution of Dirac and Weyl nodes as $J_1'/|J_1|$ is increased. The 3D plots show the spin waves for the FM configuration with $J_2$ = $J_3$ = 0 and $J_1'/|J_1|$ = (a) 0, (b) 0.5, (c)  1.0, (d)  1.5, (e) 2.0, (f)  2.5.}
  \label{fig:FMJ1P}
\end{figure}

\noindent where, in the absence of $J_2$ and $J_3$, the FM configuration is dominant for $J_1>0$ and the AFM configuration for $J_1<0$. The competition between higher-order exchange parameters produces the ZZ, DIM, and ARM phases. It should be noted that the collinear phases are not always stable without an easy-axis anisotropy.\cite{hara:09JPCM} This indicates that non-collinear phases may be the global ground state for many exchange combinations. While there may be other collinear phases that can occur, as well as non-collinear phases, due to this competition of exchange parameters, we have chosen to focus on the most likely cases.\cite{hara:09PRL,swan:09} Further consideration of other phases will be examined in future studies.

Through a comparison of the configuration energies, it is possible to create phase diagrams for various combinations of exchange parameters, which is shown in Fig. \ref{fig:PhaseStructures}(f-j). These diagrams show the general relationship of the various phases and provide an avenue for understanding the competition within the various structures.

The phase diagrams in Fig. \ref{fig:PhaseStructures}(f-j) are quite complex. Therefore, for simplification, the classical energies are normalized to $|J_1|$, where a ferromagnetic $J_1$ assumed for Fig. \ref{fig:PhaseStructures}(f) and an anti-ferromagnetic $J_1$ for Fig. \ref{fig:PhaseStructures}(g). Figures \ref{fig:PhaseStructures}(h-j) show the phase diagram iterations in which the exchange parameters are set to $0$. 

From these phase diagrams, it becomes clear that in the limit where $J_2$ = $J_3$ = 0, only the FM and AFM are present. The other collinear phases only become a stable ground state in the presence of next and next-next nearest interactions.

Furthermore, with the known regions of stability for the magnetic phases, parameters for the exchange interactions can be determined and used to construct a dynamics matrix using a similar technique to that discussed in Ref. [\onlinecite{hara:09JPCM}]. A table of the coordinates and anisotropy used are shown in table \ref{table:parameters}. Through a diagonalization of this matrix, the spin-wave dynamics can be determined.

\section{Spin-Wave Modes and Dirac Magnons}

Magnetic Dirac nodes are the crossing points of two distinct spin-wave modes; typically at a high-symmetry point. Here, the spin waves are determined for each configuration. It should be pointed out that for more complex structures, there are more SL structures. This will inherently lead to more spin-wave modes. To illustrate the spin-wave dynamics, we show the progressive evolution of the spectra as each exchange parameter ($J_1$, $J_2$, and $J_3$) is added, which is shown in a 2D plot. Here, the modes are shown along the high-symmetry pathway ($\Gamma$,M,K,$\Gamma$), as well as, in 3D plots that cover the full Brillouin zone. Table \ref{table:parameters} shows the parameters used to calculate the spin waves for each honeycomb structure, as well as the critical anisotropy term used to stabilize each collinear phase. We do find several Dirac nodes in these phases, and this is because in these configurations the SLs have inversion symmetry through specific paths and not others.

\subsection{Ferromagnetic Honeycomb Lattice}

The FM structure consists of a structural and magnetic 2-SL structure. As mentioned above, this maintains inversion symmetry and should produce magnetic Dirac bosons in the same way that one observes Dirac fermions in the electronic structure. Using the Hamiltonian discussed above, we determine the spin-wave frequencies for the FM configuration as a function of the exchange parameters to be

\begin{widetext}
\begin{equation}
\begin{array}{ll}
\omega_{FM}^{\pm} = &\pm(2{\it D}+2\,J_1+6\,{J_2}+9\,J_3+J_1^{\prime})\mp2\,J_2\,\Big(\cos \left( \sqrt {3}k_y \right) +\cos \left( \frac{3}{2}\,k_x-\frac{\sqrt {3}}{2}k_y \right) + \cos \left( \frac{3}{2}\,k_x+\frac{\sqrt {3}}{2}k_y \right)\Big) \\ &
+ \Big[ 2\,{J_1}^{2}+9\,{J_3}^{2}+{J_1^{\prime}}^{2}\pm\Big(6\, \left( J_3+J_1 \right)  \left( J_3+\frac{J_1^{\prime}}{3} \right)\Big( \cos \left( \frac{3}{2}\,k_x-\frac{\sqrt {3}}{2}k_y \right)\Big)  \\ &
+\Big(6\, \left( J_3+J_1 \right)  \left( J_3+\frac{J_1^{\prime}}{3} \right)\Big)\Big(\cos \left( \frac{3}{2}\,k_x+\frac{\sqrt {3}}{2}k_y \right)\Big)
+4\, J_3\, \left( J_1+J_3+\frac{J_1^{\prime}}{2} \right) \Big(\cos \left( \frac{3}{2}\,k_x+\frac{3\sqrt {3}}{2}k_y \right) \\&
+ \cos \left( \frac{3}{2}\,k_x-\frac{3\sqrt {3}}{2}k_y \right) + \cos \left( 3\,k_x \right)+\left( J_3+J_1 \right) \cos \left( 2\,\sqrt {3}k_y \right)\Big)+4\,{J_3}^{2}\Big(\cos \left( \frac{3}{2}\,k_x -\frac{5\sqrt {3}}{2}k_y\right)  \\& 
+\frac{1}{2}\cos \left( \frac{9}{2}\,k_x -\frac{3\sqrt {3}}{2}k_y\right) +\cos \left( \frac{9}{2}\,k_x -\frac{\sqrt {3}}{2}k_y\right) +\cos \left( \frac{9}{2}\,k_x+\frac{\sqrt {3}}{2}k_y \right) \\&
+\frac{1}{2}\cos \left( \frac{9}{2}\,k_x+\frac{3\sqrt {3}}{2}k_y \right) + \cos \left( \frac{3}{2}\,k_x+\frac{5\sqrt {3}}{2}k_y \right)+\cos \left( 3\,k_x -2\,\sqrt {3}k_y\right) +\cos \left( 3\,k_x+2\,\sqrt {3}k_y \right)\Big) \\&
+2\,{J_3}^{2}\cos \left(3\,\sqrt {3}k_y \right)+2\,J_3\, \left( J_1+2\,J_3+J_1^{\prime} \right) \Big(\cos \left( 3\,k_x -\sqrt {3}k_y\right)  + \cos \left( 3\,k_x+\sqrt {3}k_y \right)\Big) \\& 
+\left( 6\,{J_3}^{2}+ \left( 4\,J_1+4\,J_1^{\prime} \right) J_3+2\,{J_1}^{2} \right) \cos \left( \sqrt {3}k_y \right)  \Big]^{1/2},
\end{array}
\end{equation}
\end{widetext}

which consists of two distinct modes due to the 2-SL magnetic structure. Fig. \ref{fig:fmafm}(a-d) shows the spin waves of the FM configuration of the honeycomb lattice with three different exchange configurations are considered.

Fig. \ref{fig:fmafm}(b) shows a 3D representation of the spin waves considering only $J_1$ (with $J_1$ = $J_1^{\prime}$), which produces a magnetic structure that is similar to the observed and calculated electronic structure. Here, the presence of clear Dirac nodes at the $K$ and $K^{\prime}$ symmetry points is easily observed by the black line in Fig. \ref{fig:fmafm}(a). As $J_2$ is introduced (Fig. \ref{fig:fmafm}(c)), the Dirac node is shifted up in energy and the Dirac character is weakened, as observed by the loss of the linearity in the red dashed line of Fig. \ref{fig:fmafm}(a). Interestingly, the further introduction of $J_3$ in Fig. \ref{fig:fmafm}(d) shifts the modes and produces two new crossing points. One at the high-symmetry M point and one off-symmetry point that can be hard to observe in the 3D plot. Therefore, the blue dot-dashed line in Fig. \ref{fig:fmafm}(a) shows the spin-wave modes and crossover nodes.

The presence of multiple mode crossovers with $J_3$ signals to an interesting break in symmetry that introduces Dirac and Weyl points. To investigate this generation of non-symmetric modes, Fig. \ref{fig:FM-evo} shows the evolution of the spin waves for $J_1$ = $J_1^{\prime}$ = 1 with increasing $J_2$ and $J_3$. From Fig.\ref{fig:FM-evo}(a) the creation of the Weyl points is due solely to the implementation of the next-next-nearest neighbor interaction $J_3$, where Fig. \ref{fig:FM-evo}(b) shows the creation of multiple Weyl crossover points at a critical value of $J_3/J_1$ = 4/13. We can quantify the nodal pathway is dictated by 

\begin{equation}
k_y^{np} = \left\{
        \begin{array}{ll}
            \frac{4\sqrt{3}\pi}{9}, &  {\rm all} \frac{J_3}{J_1} \\
            \frac{2\sqrt {3}}{3}\cos^{-1} \left( \,{\frac {-1-\sqrt {-4J_1/J_3+13}}{4}} \right),   & \frac{4}{13}< \frac{J_3}{J_1} \\
            \frac{2\sqrt {3}}{3}\cos^{-1} \left( \,{\frac {-1+\sqrt {-4J_1/J_3+13}}{4}} \right), & \frac{4}{13}< \frac{J_3}{J_1}
        \end{array}
    \right.
\end{equation}
Furthermore, the original Dirac node at the $K$ symmetry point remains independent of the higher-order interactions.

To examine the creation of these nodes further, Figs. \ref{fig:J3e}(a-j) show the production of the nodal points in $k$-space as $J_3$ is introduced. In Fig. \ref{fig:J3e}(a), the standard six Dirac nodal points are located at $K$ and $K^{\prime}$ are present for $J_3$ = 0. As $J_3$ is increased, the sharp Dirac nodal point begins to lose its linearity (shown in Figs. \ref{fig:J3e}(b) and (c)). Once beyond the critical value of $J_3/J_1$ = 4/13, the presence of three sharp Weyl points are formed, while three more-rounded Weyl points also form at 60$^{\circ}$ angles in $k$-space (shown in \ref{fig:J3e}(d)). In Figs. \ref{fig:J3e}(e-j), the Weyl nodes moving towards the $M$ symmetry point will collide and annihilate each other leaving only the inner Weyl modes.

The annihilation of the Weyl points is likely due to the nature of the Chern number or Berry curvature for the nodes. Since they have equal and opposite Chern number, the nodal energy is increased when they come together at the $M$ point, which produces an energy gap in the spin-wave dynamics.

The change in the modes is fascinating because the Dirac node itself remains at the $K$ symmetry point. Only the created Weyl points move, which is different than the merging of Dirac point discussed by Montambaux et al. in 2009,\cite{mont:09} where it was demonstrated that the Dirac nodes could be shifted in the electronic structure with an implementation of asymmetry in the nearest-neighbor hopping parameter $t$.\cite{mont:09} The shifting of the Dirac node is due to an asymmetry placed into the interactions that increase the coupling of the two SLs and slowly shifts them towards the $M$ point. This shift can be observed in the magnetic structure as well by adding a similar asymmetry to the magnetic exchange interaction $J_1^{\prime}$.

Figure \ref{fig:FMJ1P} shows the evolution of the mode crossing in the FM configuration as $J_1^{\prime}/|J_1|$ is increased from 0 to 2.5. At $J_1^{\prime}/|J_1|$ = 0, the magnetic structure consists of 1D bands of interacting spins. Therefore, the system produces distinct Dirac lines.\cite{mull:15} As $J_1^{\prime}/|J_1|$ is increased to 1, where the standard Dirac nodes are produced. However, as $J_1^{\prime}/|J_1|>1$, the Dirac nodes shift towards the $M$ points and eventually merge in a similar manner observed for the $J_3$ modes. 

Therefore, we observe two separate phenomena. (1) the shifting of the Dirac point due to asymmetry in the exchange parameter, and (2) the production of Weyl points due to the shifting topology of the spin-wave modes with $J_3$.

\renewcommand{\arraystretch}{1.5}
\begin{table}[h!]
 \caption{A list of values used to determine the spin waves in Figures \ref{fig:fmafm} and \ref{fig:dza}.}
\centering
 \begin{tabular}{|| c | c | c ||}
 \hline
 \multicolumn{3}{||c||}{Spinwave Calculation Parameters} \\
 \hline
 Structure & $\Big(\frac{J_1}{|J_1|},\frac{J_2}{|J_1|},\frac{J_3}{|J_1|}\Big)$ & Anisotropy, $\frac{D}{J_1}$ \\
 \hline \hline
 FM & $(1,0,0)$ & $0$\\
  & $(1,1,0)$ & $0$\\
  & $(1,1,1)$ & $0$\\
 \hline
AFM & $(-1,0,0)$ & $0$\\
  & $(-1,1,0)$ & $0$\\
  & $(-1,1,-1)$ & $0$\\
 \hline
 DIM & $(-1,-8,0)$ & $3.516$\\
  & $(-1,-8,3)$ & $3.783$\\
 \hline
 ZZ & (1,-4,0) & 1.531\\
  &  (1,-4,$\frac{1}{2}$) &  1.759\\
 \hline
ARM & (1,-1,0) & 0.1250\\
  & (1,-1,-1) & 0.0396\\
 \hline
 \end{tabular}
 \label{table:parameters}
\end{table}

\subsection{Antiferromagnetic Honeycomb Lattice}

In the AFM configuration (Fig. \ref{fig:PhaseStructures}(b)), the 2-SL sites have opposite collinear spins (up and down). While this is a small and simple change, it has a dramatic effect on the modes. Using the same method as the FM configuration, the spin-wave modes can be determined analytically as

\begin{widetext}
\begin{equation}
\begin{array}{ll}
\omega_{AFM} =&\Big[2\,{{\it J_1}}^{2}+42\,{{\it J_2}}^{2}+72\,{{\it J_3}}^{2}-4\,{{\it J_3}}^{2}\Big(\cos \left(\frac{3}{2}{\it k_x}+ \frac{5\sqrt 3}{2}{\it k_y} \right) +\cos \left( \frac{3}{2}\,{\it k_x}
 -\frac{5\,\sqrt3}{2}{\it k_y}\right) \\&
 
+ \cos \left(\frac{9}{2}\,{\it k_x}+\frac{\sqrt 3}{2}\,{\it k_y} \right)+\cos \left(3\,{\it k_x}+2\,\sqrt {3}{\it k_y
}\right)+\cos \left( 3\,{\it k_x}-2\,\sqrt {3}{\it 
k_y} \right)+\cos \left( \frac{9}{2}\,{\it k_x}-\frac{\sqrt {3}}{2}\,{\it k_y} \right)\Big)  \\&

-2\,{{\it J_3}}^{2}\Big(\cos \left( \frac{9}{2}\,{\it k_x}+\frac{3\,\sqrt 3}{2}{\it k_y} \right)+\cos \left( 3\,\sqrt {3}{\it k_y} \right) +\cos \left( \frac{9}{2}\,{\it k_x}-\frac{3\sqrt {3}}{2}\,{\it k_y} \right) \Big)\\&

 + {\it J_3}\left( -36\,{\it D}+36\,{\it J_1}-108\,{\it J_2}+18\,{{\it J_1}}^{\prime} \right)+ {\it J_2}\left( 24\,{\it D}-24\,{\it J_1}-12\,{{\it J_1}}^{\prime} \right)\\&

+ {\it J_1}\left( -8\,{\it D}+4\,{{\it J_1}}^{\prime} \right) +4\,{\it D}\, \left( {\it D}-{{\it J_1}}^{\prime} \right) + \cos \left( 3\,{\it k_x} \right)\left( -4\,{{\it J_3}}^{2}+ {\it J_3}\left( -4\,{\it J_1}-2\,{{\it J_1}}^{\prime} \right) 
+4\,{{\it J_2}}^{2} \right)  \\&

+ \cos \left( \sqrt {3}{\it k_y} \right)\Big( -6\,{{\it J_3}}^{2}+ {\it J_3}\left( -4\,{\it J_1}+36\,{\it J_2}-4\,{{\it J_1}}^{\prime} \right) -20\,{{\it J_2}}^{2}+ {\it J_2}\left( -8\,{\it D}+8\,{\it J_1}+4\,{{\it J_1}}^{\prime} \right) -2\,{{\it J_1}}^{2} \Big) \\&

+ \Big(\cos \left( 3\,{\it k_x}+\sqrt {3}{\it k_y} \right)+\cos \left(3\,{\it k_x} -\sqrt {3}{\it k_y} \right)\Big)\Big( -4\,{{\it J_3}}^{2}-{\it J_3}\left(2\,{\it J1}+2\,{{\it J_1}}^{\prime} \right) +2\,{{\it J_2}}^{2} \Big) \\&

+ \Big(\cos \left( \frac{3}{2}\,{\it k_x}+\frac{3\sqrt {3}}{2}\,{\it k_y} \right)+\cos \left( \frac{3}{2}\,{\it k_x}-\frac{3\sqrt {3}}{2}\,{\it k_y}\right)\Big)\Big( -4\,{{\it J_3}}^{2}- {\it J_3}\left( 4\,{\it J_1}+2\,{{\it J_1}}^{\prime} \right) +4\,{{\it J_2}}^{2} \Big)  \\&

+ \Big(\cos \left( \frac{3}{2}\,{\it k_x}-\frac{\sqrt {3}}{2}\,{\it k_y} \right)+\cos \left( \frac{3}{2}\,{\it k_x}+\frac{\sqrt {3}}{2}\,{\it k_y} \right)\Big) \Big(-6\,{{\it J_3}}^{2}-{\it J_3}\left(6\,{\it J_1}-36\,{\it J_2}+2\,{{\it J_1}}^{\prime} \right) \\&

-20\,{{\it J_2}}^{2}+  {\it J_2}\left( -8\,{\it D}+8\,{\it J_1}+4\,{{\it J_1}}^{\prime} \right)-2\,{\it J_1}\,{{\it J_1}}^{\prime} \Big) +\cos \left( 2\,\sqrt {3}{\it k_y} \right) \left( -4\,{\it J_1}\,{\it J_3}+2\,{{\it J_2}}^{2}-4\,{{\it J_3}}^{2} \right) \Big]^{1/2},
\end{array}
\end{equation}
\end{widetext}
As shown in Fig. \ref{fig:fmafm}(e), the AFM configuration does not produce two modes. There is a single Goldstone mode at the $\Gamma$ point. The reduction in modes is due to the flip of spin between the structural sublattices reduces the system to a 1-SL system. Therefore, the AFM configuration will only produce a single spin-wave mode or more precisely double degenerate modes. Furthermore, these modes are not affected by the addition of higher-order interactions ($J_2$ and $J_3$) and the introduction of anisotropy destroys merely the Goldstone mode.

The loss of Dirac symmetry becomes clear when only the spin interactions are considered, such as in Fig. \ref{fig:inversion}(e). Geometrically, the structure remains chiral, however magnetically, the AFM configuration loses its inversion symmetry, and SL2 becomes similar to SL1, without regard to the geometric position. The energy for each sublattice is identical, creating degenerate modes in this phase and therefore only showing one mode. Therefore, there are no Dirac modes in AFM configuration regardless of interactions.

In comparison to the FM configuration, the magnetic environment at site 2 is a simple flip of inversion symmetry from site 1 (as shown in Fig. \ref{fig:inversion}(c)). However, in the AFM configuration, the magnetic environment at site 2 is equal to an inversion of the site 1. It is a 180$^{\circ}$ rotation about the dashed line shown in Fig. \ref{fig:inversion}(e). This break in inversion symmetry reduces the system to a single SL system and eliminates the Dirac nodes that are present in the FM phases.

\begin{figure}
  \centering
  \includegraphics[width=3.25in]{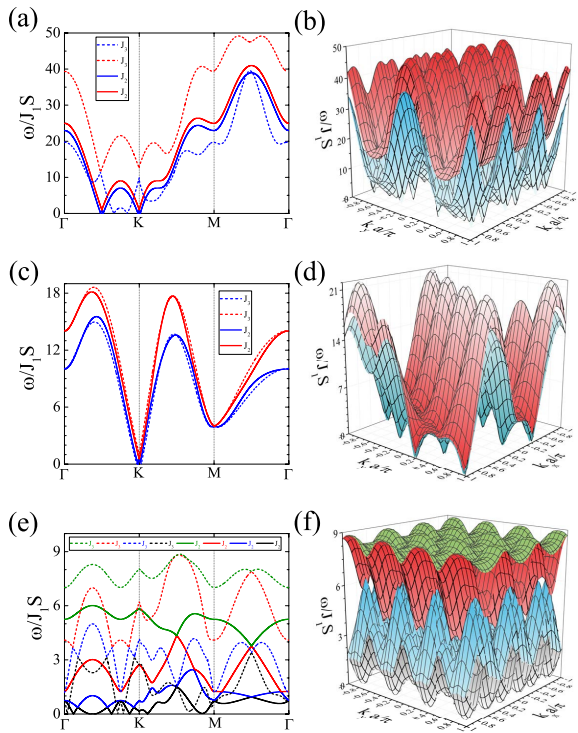}
  \caption{To the left, the 2-D spin waves of a DIM, ZZ, and ARM honeycomb structures with added levels of exchange-interactions. Here only the $J_2$ and $J_3$ spin exchange interactions have been mapped due to the fact that at $J_2=0$ and $J_3=0$ only a FM or an AFM structure is left. To the right, the 3-D spin waves of the DIM, ZZ, and ARM honeycomb structures with all three spin exchange interactions, as showing the other spin exchange interactions is no more conducive to their understanding. The parameters are given in table \ref{table:parameters}. }
  \label{fig:dza}
\end{figure}

\subsection{Other Antiferromagnetic Honeycomb Lattices}

To examine the other AFM phases, the spin model requires next and next-next nearest neighbor interactions. Furthermore, we need the introduction of easy-axis anisotropy in order stabilize the phases\cite{swan:09}. The need for anisotropy reveals the presence of non-collinear phases below the critical anisotropy. While we do not investigate the non-collinear phases in this study, the ordering wave-vectors for the non-collinear spin-waves can be determined by the $k$ points of the Goldstone modes that are present at the critical anisotropy values\cite{hara:09PRL}.

From Fig.\ref{fig:PhaseStructures}, the dimerized (DIM) configuration consists of spin up-up and down-down dimers, while the zig-zag (ZZ) and armchair (ARM) configurations include of alternating stripes of up and down spins along the zig-zag and armchair directions, respectively.

The DIM, ZZ, and ARM configurations complicate the Hamiltonian matrix due to the increase in the magnetic sublattices. The DIM and ZZ structures produce a four sub-lattice (4-SL) magnetic system, while the ARM structure produces an eight sub-lattice (8-SL) magnetic system. Because of the complicated magnetic structures, analytical solutions for the spin-waves were not able to be obtained. However, the spin-wave dynamics were calculated numerically.

Figure \ref{fig:dza} shows the calculated spin-wave spectra for the DIM, ZZ, and ARM configurations. These do indicate that you may obtain Weyl nodes depending on the specific configuration and the included interactions. However, it is clear from the broken inversion symmetries in the magnetic structures that you should not produce Dirac nodes, which is consistent with the sample spectra that are provided.

Due to their complicated magnetic structures, inversion symmetry is not consistently held through all three structures. For the DIM and ZZ configurations, inversion symmetry is only held between two nearest neighbor spins, which also reduces the number of modes to two. While the ARM configuration has inversion symmetry on downward facing like spins, the number of modes is reduced to four. 

Understanding these more complicated magnetic structures and their underlying non-collinear states can help lead to further insight into the character of Dirac and Weyl modes and materials.

\section{Conclusion}

In this study, we examine the creation and evolution of Dirac and Weyl nodes in the magnetic honeycomb lattice. Using linear spin-wave theory, we determined the spin dynamics and phases diagrams for five common collinear phases.

In the AFM configuration, the breaking of inversion symmetry in the magnetic structure eliminates the presence of a Dirac node. However, in more complex AFM configurations (DIM, ZZ, and ARM), the crossing of spin waves can produce other Dirac-like nodes. It should be noted that only selected collinear phases have been considered. The more complicated AFM configurations are only stable with easy-axis anisotropy, which indicates the presence of non-collinear phases below that critical anisotropy. In the future, we plan to investigate the non-collinear phases further using Monte Carlo simulations, which is currently out of the scope of this study.

With regards to the FM configuration, we observe two main phenomena. The introduction of asymmetry in the honeycomb lattice can shift the Dirac nodes in $k$-space, while additional higher-order interactions can produce multiple Weyl nodes from the $K$ and $K^{\prime}$ symmetry points. This ability to control the creation of Dirac and Weyl nodes through variations in exchange interactions leads to the possibility of topological manipulation through the use of external fields in the time domain. Since magnetic superexchange is controlled through the nature of orbital overlap, the presence of superficial fields can affect exchange pathways in the time domain through dynamic interactions in the topology.

\section*{Acknowledgements}

D.B. and J.T.H. acknowledges support by the Institute for Materials Science at Los Alamos National Laboratory. Furthermore, travel support was provided, in part, by the National Science Foundation under Grant No. NSF PHY-1125915 (J.T.H.). The work at Los Alamos National Laboratory was carried out under the auspice of the U.S. DOE and NNSA under Contract No. DEAC52-06NA25396 and supported by U.S. DOE Basic Energy Sciences Office (A.V.B).

\end{document}